\documentclass[twocolumn,showkeys,aps,prl,showpacs]{revtex4-1}
\usepackage{graphicx}
\usepackage[CJKbookmarks,dvipdfm,colorlinks,linkcolor=blue,citecolor=blue]{hyperref}

\begin{document}

\title{Importance of spin-orbit coupling  in  power factor calculations for  half-Heusler $\mathrm{ANiB}$ (A=Ti, Hf, Sc, Y; B=Sn, Sb, Bi) }

\author{San-Dong Guo}
\affiliation{Department of Physics, School of Sciences, China University of Mining and
Technology, Xuzhou 221116, Jiangsu, China}
\begin{abstract}
We investigate  the spin-orbit coupling (SOC) effects on the electronic structures and  semi-classic transport coefficients
 of half-Heusler $\mathrm{ANiB}$ (A=Ti, Hf, Sc, Y; B=Sn, Sb, Bi)  by using  generalized gradient approximation (GGA). Calculated results show that SOC splits the valence bands at high symmetry $\Gamma$ point, and modifies the outline of $\Gamma$-centered valence bands, which has remarkable effects on the electron transport properties.
 Thermoelectric properties are performed through solving Boltzmann transport equations within the constant scattering time approximation.  It is found that the  compounds containing Sn atom  have larger power factor in p-type doping than ones in n-type doping, and it is just the opposite for compounds containing Sb and Bi elements.
 The SOC has  obvious detrimental influence on power factor in p-type doping, while has a negligible effect in n-type doping. These can be understood by considering the effects of SOC on the valence bands and conduction bands. The maximum power factors (MPF)  are extracted in n-type and p-type doping with GGA and GGA+SOC, and the MPF at 300 K with SOC is predicted to be  about 4.25\%$\sim$44.13\%  smaller than that without SOC in the case of p-type doping  for $\mathrm{ANiB}$ (A=Ti, Hf, Sc, Y; B=Sn, Sb, Bi).  Therefore,  it is crucial to consider SOC effects
 for theoretical analysis in the case of p-type doping in half-Heusler compounds composed of heavy elements.

\end{abstract}

\pacs{72.15.Jf, 71.20.-b, 71.70.Ej, 79.10.-n~~~~~~~~~~~~~~~~~~~~~~~~guosd@cumt.edu.cn}

\maketitle

\section{Introduction}
Thermoelectric devices are potential energy converters to solve energy problems, which can convert waste heat directly to
electricity using the Seebeck effect. The  performance  of thermoelectric material is characterized by the dimensionless thermoelectric figure of merit\cite{s1,s2}, $ZT=S^2\sigma T/(\kappa_e+\kappa_L)$, where S, $\sigma$, T, $\kappa_e$ and $\kappa_L$ are the Seebeck coefficient, electrical conductivity, absolute working temperature, the electronic and lattice thermal conductivities, respectively. Many materials have been
identified for thermoelectric applications, such as bismuth-tellurium systems\cite{s3,s4}, silicon-germanium alloys\cite{s5,s6}, lead chalcogenides\cite{s7,s8} and skutterudites\cite{s9,s10}. Heusler compounds have wide applications in spintronics, shape memory alloys, superconductors, topological insulators and thermoelectrics\cite{s11}, and half-Heusler  have attracted intensive research interest  as moderate temperature thermoelectric materials due to being environmentally friendly, mechanically
and thermally robust\cite{s12,s13,s14,s15,s16,s17,s18}.

Recently, the transport properties of materials can be calculated accurately by combining the first
principles band structure calculations and the Boltzmann transport theory\cite{b0,b01,b}.
Many theoretical simulation calculations have been performed for  thermoelectric properties of half-Heusler\cite{t1,t2,t3,t4,t5,t6}.
However, most of them do not consider the SOC effects on transport properties. As is well known, spin-orbit interaction plays a key role in materials composed of heavy elements such as Bi or Sb, and SOC can induce topological insulators\cite{t61}.
In ref.\cite{e1},  relativistic effects in  thermopower calculations
for $\mathrm{Mg_2X}$ (X=Si, Ge, Sn) are very remarkable, and  have a  detrimental influence on the thermoelectric performance of p-type $\mathrm{Mg_2X}$. So, it is very  necessary to know the SOC effects on the  thermoelectric properties of half-Heusler compounds containing heavy elements.

Here, we investigate  SOC effects on  thermoelectric properties  of half-Heusler $\mathrm{ANiB}$ (A=Ti, Hf, Sc, Y; B=Sn, Sb, Bi). It is found that SOC strongly affects the top valence bands near $\Gamma$ point, which leads to a detrimental effect
on p-type power factor. On the other hand, the SOC influence  on
conduction bands near the Fermi level  is little,  and a negligible SOC  effect  on n-type power factor is observed.
At the presence of SOC, the MPF at 300 K is predicted to be  about up to  44.13\%  smaller than that without SOC in the case of p-type doping  for YNiBi. So, SOC is vital for the  thermoelectric properties  of half-Heusler compounds containing heavy element.

\begin{figure}
  \includegraphics[width=7.6cm]{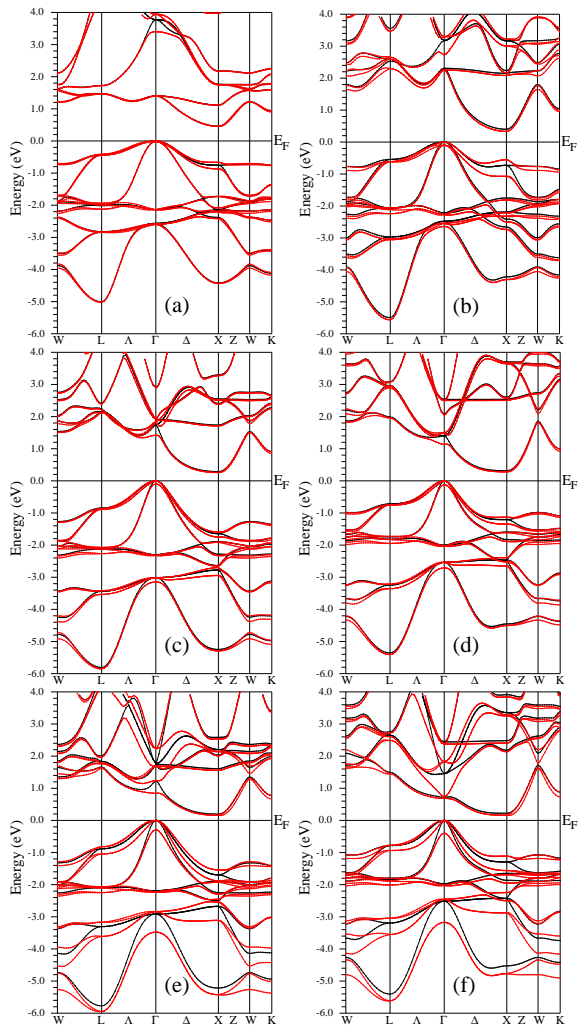}
  \caption{(Color online) The energy band structures  by using GGA (Black lines) and GGA+SOC (Red lines) (a) TiNiSn (b)HfNiSn (c) ScNiSb (d) YNiSb (e) ScNiBi (f) YNiBi.}\label{band}
\end{figure}

The rest of the paper is organized as follows. In the next
section, we shall give our computational details. In the third section, we shall present our main calculated results and
analysis. Finally, we shall give our conclusion in the fourth
section.

\section{Computational detail}
We use a full-potential linearized augmented-plane-waves method
within the density functional theory (DFT) \cite{1}, as implemented in
the package WIEN2k \cite{2}.  We use the popular GGA\cite{pbe} for the
exchange-correlation potential  to do our DFT
calculations. The full relativistic effects are calculated
with the Dirac equations for core states, and the scalar
relativistic approximation is used for valence states
\cite{10,11,12}.  The SOC was included self-consistently
by solving the radial Dirac equation for the core electrons
and evaluated by the second-variation method\cite{so}. We use 5000 k-points in the
first Brillouin zone for the self-consistent calculation. We make harmonic expansion up to $\mathrm{l_{max} =10}$ in each of the atomic spheres, and set $\mathrm{R_{mt}\times k_{max} = 8}$. The self-consistent calculations are
considered to be converged when the integration of the absolute
charge-density difference between the input and output electron
density is less than $0.0001|e|$ per formula unit, where $e$ is
the electron charge. Transport calculations
are performed through solving Boltzmann
transport equations within the constant
scattering time approximation as implemented in
BoltzTrap\cite{b}, which has been applied successfully to several
materials\cite{b1,b2,gsd}. To
obtain accurate transport coefficients, we use 200000 k-points in the
first Brillouin zone for the energy band calculation.
\begin{figure}
  \includegraphics[width=7cm]{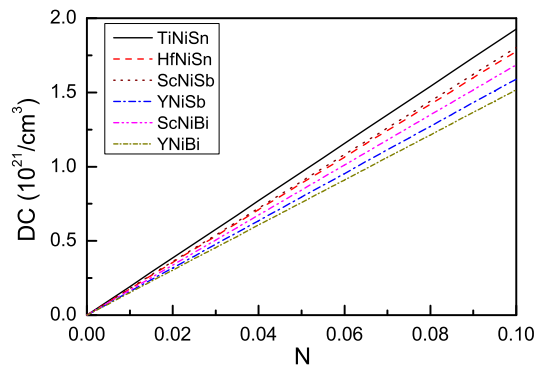}\\
  \caption{(Color online) The corresponding relation between the doping concentration and electrons  or holes  per unit cell. }\label{tu1}
\end{figure}

\begin{table}[!htb]
\centering \caption{ The experimental lattice constant $a$  ($\mathrm{{\AA}}$); the calculated gap values  with GGA $E_1$ (eV) and GGA+SOC $E_2$ (eV); $E_1-E_2$ (eV);  spin-orbit splitting $\Delta$ (eV)  at the ${\Gamma}$ point near the Fermi level in the valence bands. These values in the parentheses are GGA gaps in ref.\cite{t5}.}\label{tab}
  \begin{tabular*}{0.48\textwidth}{@{\extracolsep{\fill}}cccccc}
  \hline\hline

 Name & $a$  & $E_1$ & $E_2$&$E_1-E_2$ &$\Delta$\\\hline\hline
 TiNiSn &5.921  & 0.460 (0.451) &0.452&0.008 &0.025\\\hline
 HfNiSn&6.084  & 0.401 (0.396) & 0.333& 0.068&0.100\\\hline
 ScNiSb &6.055  & 0.286 (0.281) &0.259&0.027& 0.095\\\hline
 YNiSb & 6.312  & 0.314 (0.311) &0.281&0.033 &0.117\\\hline
 ScNiBi&6.191  & 0.195 (0.191) & 0.154& 0.041&0.291\\\hline
 YNiBi &6.411  & 0.221 (0.219) &0.156& 0.065&0.398\\\hline\hline
\end{tabular*}
\end{table}
\section{MAIN CALCULATED RESULTS AND ANALYSIS}
\begin{figure*}
  \includegraphics[width=15.5cm]{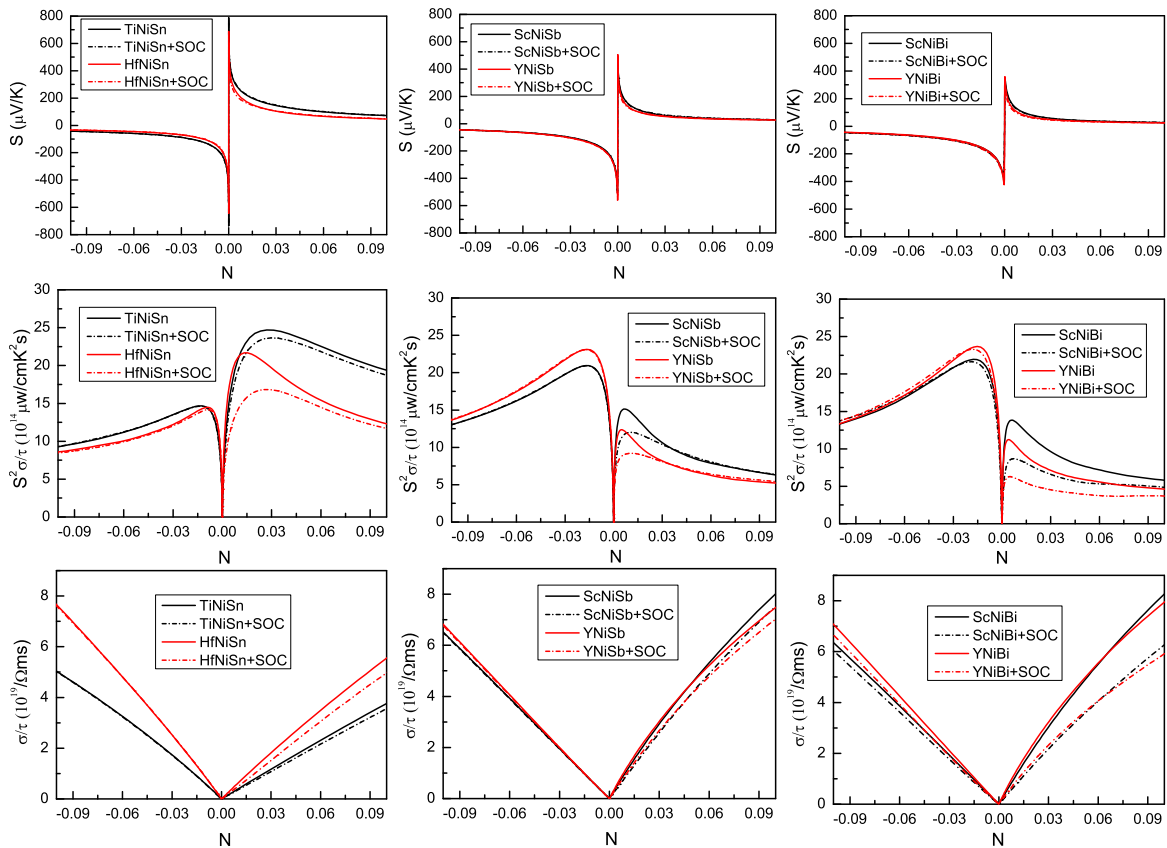}
  \caption{(Color online) At temperature of 300 K,  transport coefficients  as a function of doping levels (electrons [minus value] or holes [positive value] per unit cell):  Seebeck coefficient S (Top panel),  power factor with respect to scattering time $\mathrm{S^2\sigma/\tau}$ (Middle panel) and electrical conductivity with respect to scattering time  $\mathrm{\sigma/\tau}$ (Bottom panel) calculated with GGA (Solid line) and GGA+SOC (Dotted line). }\label{eb}
\end{figure*}

Half-Heusler $\mathrm{ANiB}$ (A=Ti, Hf, Sc, Y; B=Sn, Sb, Bi) forms a MgAgAs type
of structure with space group $F\bar{4}3m$, where A, Ni, and B atoms occupy
Wyckoff positions 4a (0, 0, 0), 4c (1/4, 1/4,
1/4), and 4b (1/2, 1/2, 1/2) positions, respectively. The experimental lattice crystal structures\cite{le} are used to do our calculations, and the lattice constants $a$ are listed in \autoref{tab}. Here, we investigate the electronic structures of  TiNiSn, HfNiSn, ScNiSb, YNiSb, ScNiBi and YNiBi with the 18 valence electron count (VEC) per unit cell by using GGA and GGA+SOC, and present their energy band structures in \autoref{band}. They are all indirect-gap semiconductors, with the conduction band minimum (CBM) at high symmetry point X and  valence band maximum (VBM) at the $\Gamma$ point. These gaps are produced due to the strong hybridization of d states of the A and Ni atoms. The valence bands near the Fermi level are dominated by the
A-d state hybridized with the Ni-d and B-p states, while the bottom of the conduction bands are constructed  mostly  by A-d and Ni-d states. Based on our calculation, the GGA gap values vary from about 0.195 eV to 0.460 eV, and GGA+SOC ones change from 0.154 eV to 0.452 eV.  Our GGA gap values are well consistent with other theoretical values\cite{t5} calculated by using density functional projector augmented plane-wave method within the GGA.
This SOC effect on gap strongly depends on A and B atoms, and the larger gap reduce means  more obvious influence on conduction bands near Fermi level. We show the related gap values in \autoref{tab}. To describe the SOC effects on the valence bands near Fermi level,  spin-orbit splitting at the ${\Gamma}$ point near the Fermi level in the valence bands are calculated, and are listed in \autoref{tab}. These data show SOC has larger influence on the valence bands with respect to conduction bands.

\begin{figure*}
  \includegraphics[width=15.5cm]{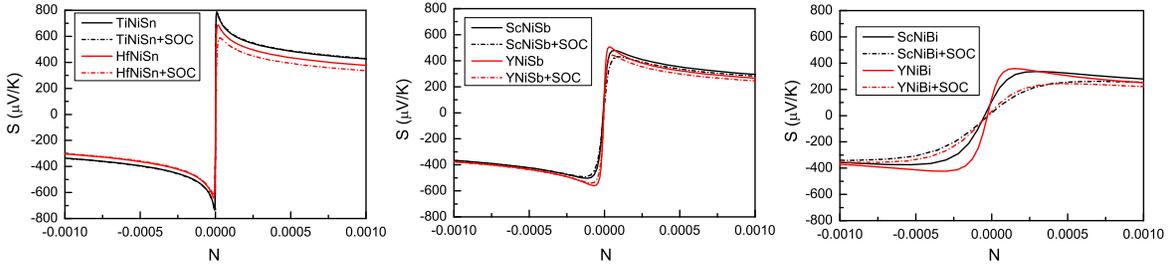}
  \caption{(Color online) At temperature of 300 K,  Seebeck coefficient S (enlarged near energy gap)  as a function of doping levels (electrons [minus value] or holes [positive value] per unit cell)   calculated with GGA (Solid line) and GGA+SOC (Dotted line). }\label{eb1}
\end{figure*}

Half-Heusler is  considered as a kind of  potential  thermoelectric material for converting  heat directly to
electricity.  The calculations of the semi-classic transport coefficients as a function
of doping level are performed  within constant scattering time approximation Boltzmann theory, and   the temperature and doping dependence of the band structure are supposed to have  a negligible  effect on the transport coefficients.
First, we plot the corresponding relation between the doping concentration and electrons  or holes  per unit cell in \autoref{tu1}, and it is natural that they have a linear relation.
\autoref{eb}  shows the Seebeck coefficient S,  power factor with respect to scattering time $\mathrm{S^2\sigma/\tau}$ and electrical conductivity with respect to scattering time  $\mathrm{\sigma/\tau}$ as  a function of doping levels  at the temperature of 300 K by using GGA and GGA+SOC.  To clearly see the difference of Seebeck coefficient between GGA and GGA+SOC, the enlargers  near the gap are present in \autoref{eb1}.  The negative doping levels  imply the
n-type doping with the negative Seebeck coefficient, and
the positive doping levels mean p-type doping with the positive Seebeck coefficient.
\begin{figure}
  \includegraphics[width=6.8cm]{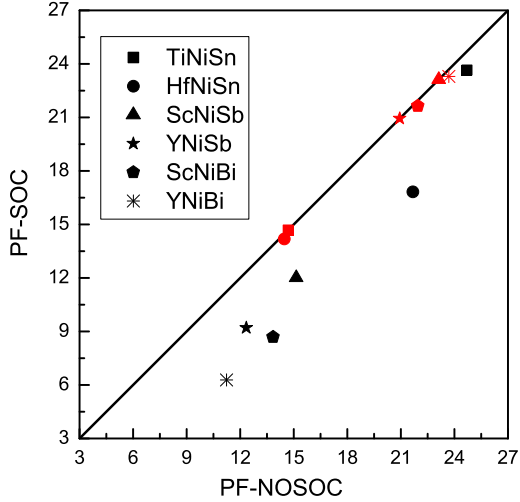}\\
  \caption{(Color online) At temperature of 300 K, the maximal power factor for p-type doping (Black mark) and n-type doping (Red mark).  The horizontal axis represents GGA values, and the vertical axis shows  GGA+SOC values.}\label{tu2}
\end{figure}

It is clearly seen that these  compounds containing Sn have larger Seebeck coefficient in p-type doping than in n-type doping, while it is totally contrary to those  compounds containing Sb and Bi. Due to S being proportional to effective mass $m^\ast$,
the analysis of the effective mass can explain the differences.  From \autoref{band}, we can see that the VBM have larger effective mass than CBM for TiNiSn and HfNiSn, while the CBM of  compounds containing Sb and Bi have larger one than VBM.
(The CBM has a dominant contribution to thermoelectric properties in the case of n-type doping, while VBM for p-type doping.)
Power factor with respect to scattering time $\mathrm{S^2\sigma/\tau}$ changes in the same trend with Seebeck coefficient, and electrical conductivity with respect to scattering time  $\mathrm{\sigma/\tau}$ changes in the opposite trend simultaneously.

Calculated results show a negligible  SOC effect  on S,  $\mathrm{S^2\sigma/\tau}$ and  $\mathrm{\sigma/\tau}$ in n-type doping for the compounds containing Sn and Sb atoms.  However, in p-type doping, a detrimental influence of SOC on the thermoelectric performance of the six kinds of compounds except the S of TiNiSn is observed. These can be understood by that SOC has larger effects on the valence bands near the Fermi level than the conduction bands. For ScNiBi and YNiBi, SOC has a observable detrimental influence on their thermoelectric performance in n-type doping, which is because they contain heavier Bi atom, although the Bi atom contributes  a small weight to  the conduction bands near the Fermi level. Finally, the maximum power factors (MPF) in unit of $\tau\times10^{14}$$\mu W/(cm K^2 s)$ are extracted in n-type and p-type doping with GGA and GGA+SOC, and are plotted in \autoref{tu2}. It is obvious that SOC has little influences on n-type MPF, and has remarkable effects on p-type MPF.
At temperature of 300 K, p-type TiNiSn have the largest power factor, followed by  n-type YNiBi and ScNiSb.
We  summarize related MPF calculated with GGA and GGA+SOC  in p-type and n-type doping, the  corresponding
 doping levels and other theoretical values in \autoref{tab1}.
Our GGA MPF and the corresponding doping levels agree well with other calculated values\cite{t5}.
\begin{table*}[!htb]
\centering \caption{The MPF in unit of $\tau\times10^{14}$$\mu W/(cm K^2 s)$ and the corresponding
 doping levels (electrons [n-type]  or holes [p-type] per unit cell)  by using GGA and GGA+SOC. These values in the parentheses are GGA MPF in ref.\cite{t5}.}\label{tab1}
  \begin{tabular*}{0.96\textwidth}{@{\extracolsep{\fill}}cc|cccccc}
  \hline\hline
   &        & TiNiSn          &HfNiSn          & ScNiSb               &YNiSb      &ScNiBi   &YNiBi \\\hline
GGA&MPF(p)  &24.70 (25.13)   &21.67 (22.03)    & 15.14 (15.82)   &12.35 (13.18)     &13.84 (14.54)   &11.24 (11.98) \\
   &p doping&0.027 (0.028)   &0.015 (0.014)    & 0.007 (0.006)   &0.005 (0.005)     &0.006 (0.006)   &0.004 (0.004) \\
   &MPF(n)  &14.68 (14.45)  &14.47  (14.44)     & 23.13 (20.64)   &20.93 (22.83)    &21.95 (21.63)   &23.68 (23.44) \\
   &n doping&0.013 (0.013)  &0.009 (0.009)     & 0.017  (0.016)  &0.015 (0.017)     &0.018 (0.016)   &0.015 (0.015) \\  \hline\hline
GGA+SOC&MPF(p)  &23.65    &16.83              & 12.02             &9.20             &8.69             &6.28 \\
   &p doping&0.030         &0.028              & 0.010              &0.011          &0.007           &0.005 \\
   &MPF(n)  &14.68         &14.18              & 23.13             &20.93           &21.64             &23.30 \\
   &n doping&  0.013      &0.008               & 0.017              &0.015          &0.018              &0.019 \\ \hline\hline

\end{tabular*}
\end{table*}

\section{Discussions and Conclusion}

The SOC removes the band degeneracy, and the importance of SOC  gradually
increases with increasing atomic number of A and B atoms. The valence bands around the high symmetry  $\Gamma$ point can show
obvious relativistic effects.  The spin-orbit splitting  removes  the
degeneracy of electronic states at $\Gamma$ point,  and modifies the outline of bands, which produces the remarkable effects on the p-type power factor. According to the spin-orbit splitting $\Delta$ in the \autoref{tab} and the MPF in \autoref{tab1}, it is found that the larger $\Delta$ leads to the more obvious detrimental influence on p-type MPF. The MPF by using GGA+SOC in p-type doping
is about 4.25\%, 20.60\%, 22.33\%, 25.50\%, 37.21\% and 44.13\% smaller than that with GGA  for TiNiSn, ScNiSb, HfNiSn, YNiSb, ScNiBi and YNiBi, respectively,  and their related spin-orbit splitting $\Delta$ also gradually increases from 0.025 eV to 0.398 eV. In fact, the power factor decay in n-type doping is connected to  difference value between gap with GGA and gap with GGA+SOC ($E_1-E_2$), and the larger gap difference value induces the larger decay.

In summary, GGA and GGA+SOC are  chosen to investigate electronic structures and thermoelectric properties  of half-Heusler $\mathrm{ANiB}$ (A=Ti, Hf, Sc, Y; B=Sn, Sb, Bi).  The strength of  SOC influences on valence and conduction bands near the Fermi level is shown by the related gaps  $\Delta$ and $E_1-E_2$.
It is found that the power factors of  p-doped TiNiSn and HfNiSn are
much higher than the values attained in n-type doping, and it is opposite to half-Heusler compounds containing Sb and Bi elements. Calculated results show that the SOC is a significant factor decreasing
the power factor of p-type $\mathrm{ANiB}$,  especially for ScNiBi and YNiBi.
In p-type doping, the MPF  reduce of YNiBi by using GGA+SOC is as much as 44.13\% with respect to GGA.
So, it is very necessary for power factor calculations  to consider SOC, when half-Heusler compounds are composed of heavy elements such as Bi or Sb.

\begin{acknowledgments}
This work is supported by the National Natural Science Foundation of China (Grant No. 11404391). We are grateful to the Advanced Analysis and Computation Center of CUMT for the award of CPU hours to accomplish this work.
\end{acknowledgments}

\end{document}